\documentclass{article}

\begin{document}

\begin{center}
\large \bf COLD DARK MATTER AND PRIMORDIAL\\[2mm]
SUPERHEAVY PARTICLES
\end{center}
\vspace{3mm}
\begin{center}
 A. A. GRIB\footnote{
 E-mail: grib@friedman.usr.lgu.spb.su } \\[2mm]

{\small A.Friedmann Laboratory for Theoretical Physics,  \\
30/32 Griboedov can, St.Petersburg, 191023, Russia }
\end{center}

\begin{center}
 YU. V. PAVLOV \\[2mm]  

{\small Institute of Mechanical Engineering,
Russian Academy of Sciences, \\
 61 Bolshoy, V.O., St.Petersburg, 199178, Russia}
\end{center}

\begin{abstract}
\noindent
The hypothesis that cold dark matter consists of primordial
superheavy particles, the decay of short lifetime component of
which led to the observable mass of matter while long living
component survived up to modern times manifesting its
presence in high energetic cosmic rays particles is investigated.
\end{abstract}

\section{Introduction}

  In this paper we shall give some new evaluation of our
proposal made in papers~\cite{GrPv},~\cite{GrPv2}
according to which experimental
observations of high energetic cosmic particles with the energy
higher than the Greizen--Zatsepin--Kuzmin limit~\cite{GZK} are
interpreted as decays of superheavy particles forming  the cold
dark matter in the vicinity of our Galaxy.
    As it is known~\cite{GD}
creation of superheavy particles with the mass of the order of
the Grand Unification scale with the subsequent decays of its
short living component on quarks and leptons with baryon
charge and $CP$-nonconservation can lead to explanation of the
observable matter.
    In Refs.~\cite{GrPv},~\cite{GrPv2} the idea that long
living component of
primordial superheavy particles can survive up to modern time
in the form of cold dark matter was discussed and some rough
estimates of its density were made.

      Different  experimental collaborations --- AGASA~\cite{Takeda},
Haverah Park~\cite{LRW},  Fly Eye~\cite{Bird},  Yakutsk~\cite{Yak}
observe in cosmic rays events corresponding to the  energy higher
than $10^{20}$\,eV.
    According to the standard theory of cosmic rays if the most energetic
particles  come from  the other galaxies accelerated by the
magnetic fields,  protons and neutrons must interact  during the
flight with photons of the primordial background radiation.
    This radiation in the reference frame of cosmic particles moving
with big velocity interacts so that pions are created.
    So cosmic particles must be decelerated and the
Greizen--Zatsepin--Kuzmin cutoff of the high energy tail of the
spectrum  of cosmic rays evaluated as $10^{20}$\,eV is predicted.
    Many different hypotheses to explain the observable
events were presented, one of them being the existence of
superheavy massive neutral particles of cold dark matter.

     Here we shall discuss just this hypothesis.
The reason for this is to consider these particles as relics of
creation of particles in the early Friedmann Universe which resulted
in the origination of visible matter, i.e. in the Eddington
number of protons and leptons.
    Strong gravitational  field of the early expanding Friedmann
Universe created from vacuum pairs of some $X$ and anti-$X$ particles
of the mass of the order of the Grand Unification order at the
Compton time from the singularity.
    Due to nonconservation of the baryon charge in
some model of Grand Unification which is not specified in our
paper these particles being created by gravity with the definite
baryon charge then decay similarly to neutral $K$-mesons as some
short living and long living components.
    Our idea is to look for  such values of the parameters of
the model which can lead to the cosmological life-time of the
long living component so that it can exist today while the
short living component decayed close to the Grand Unification time.
    Our calculations of particle creation in the early Friedmann
Universe~\cite{GMM} give the Eddington number of superheavy
particles created by gravity.
    If these particles did not decay they would lead to a quick
collapse of the closed Friedmann Universe or to a totally different
from  the observed open or quasi-Euclidean Universe.
    So these particles must decay on light particles
(here we suppose the energetic desert hypothesis).
    But these decays are different for short living and
long living components.
    Short living components decay in the time when the Grand Unification
symmetry  is not broken while the long living component survives
the symmetry breaking.
    However the number of long living particles is not equal to
the  number of created superheavy particles because similar to
the behaviour of neutral $K$-mesons created by strong interactions
and decaying through weak interactions if long living particles decay
not in  vacuum but in the substance with nonzero baryon charge they
will be converted in the short living particles and will quickly  decay.
    This process will be dependent on the density of particles.
This density is high in the early Universe and is small at the modern era.
    Now proceed to the model.

\section{Model and Numerical Estimates}

   At first suppose superheavy $X$-particles to be scalar particles.
     Total number of massive scalar particles created in
Friedmann radiation dominated Universe
(scale factor $a(t)=a_0\, t^{1/2}$)      inside the horizon is as it
is known~\cite{GMM}:
    \begin{equation}
N=n^{(0)}(t)\,a^3(t)=b^{(0)}\,M^{3/2}\,a_0^3 \ ,
\label{NbM}
\end{equation}
   where $b^{(0)} \approx 5.3 \cdot 10^{-4}$
($ N \sim 10^{80} $ for $ M \sim 10^{15} $\,Gev, see Ref.~\cite{GMM}).
    For the time ${t \gg M^{-1}} $ there is an era of going from the
radiation dominated model to the dust model of superheavy particles
    \begin{equation}
t_X\approx \left(\frac{3}{64 \pi \, b^{(0)}}\right)^2
\left(\frac{M_{Pl}}{M}\right)^4 \frac{1}{M}  \,.
\end{equation}
    If $M \sim 10^{14} $\,Gev,  $\ t_X \sim 10^{-15} $\,sec, if
$M \sim 10^{13} $\,Gev --- $t_X \sim 10^{-10} $\,sec.
    So the life time of short living $X$-mesons must be smaller then
$t_X $\,.

   Let us define $d $ --- the permitted part of long living
$X$-mesons --- from the condition: on the moment of
recombination $t_{rec} $ in the observable Universe one has
$
d\,\varepsilon_X(t_{rec}) =\varepsilon_{crit}(t_{rec})  \,,
$
where $\varepsilon_{crit}$ is the critical density for the time $t_{rec}$.
    It leads to
\begin{equation}
d=\frac{3}{64 \pi \, b^{(0)}}\left(\frac{M_{Pl}}{M}\right)^2
\frac{1}{\sqrt{M\,t_{rec}}}\, .
\label{d}
\end{equation}
   For $M=10^{13} - 10^{14} $\,Gev one has
$d \approx 10^{-12} - 10^{-14} $\,.
    Using the estimate for the velocity of change of the concentration of
long living superheavy particles~\cite{BBV}
$|\dot{n}_x| \sim 10^{-42}\, \mbox{cm}^{-3}\,\mbox{sec}^{-1} $,
and taking the life time $\tau_l $ of long living particles as
$2\cdot 10^{22} $\,sec, we obtain concentration
$n_X \approx 2\cdot 10^{-20} \,\mbox{cm}^{-3} $ at the modern epoch,
corresponding to the critical density for $M=10^{14} $\,Gev\,.

    Now let us construct the toy model which can give: \ \
a) short living $X$-mesons decay in time
   $\tau_q < 10^{-15} $\,sec, (more wishful is
   $\tau_q \sim 10^{-38} - 10^{-35} $\,sec),
   long living mesons decay with $\tau_l > t_U \approx 10^{18}$\,sec
   \  ($t_U $ is the age of the Universe),         \ \
b) one has small $ d \sim 10^{-14} - 10^{-12} $ part of long living
   $X$-mesons, forming the dark matter.

   Baryon charge nonconservation with $CP$-nonconservation in full
analogy with the $K^0$-meson theory with nonconserved hypercharge and
$CP$-nonconserva\-tion leads to the effective Hamiltonian of the decaying
$X, \bar{X}$ - mesons with nonhermitean matrix.

  For the matrix of the effective Hamiltonian
$ H=\{ H_{ij} \}, \ {i,j=1,2}$  let $H_{11}\! =\! H_{22}$
due to $CPT$-invariance.
    Denote
$\ \varepsilon=(\sqrt{\vphantom{ }H_{12}} - \sqrt{H_{21}}\,)\, / \,
(\sqrt{H_{12}} + \sqrt{H_{21}} \, )$.
    The eigenvalues $\lambda_{1,2} $ and eigenvectors
$|\Psi_{1,2}\rangle $  of matrix $H$ are
    \begin{equation}
\lambda_{1,2} = H_{11} \pm \frac{H_{12}+H_{21}}{2} \,
\frac{1-\varepsilon^2}{1+\varepsilon^2} \,,
\end{equation}
    \begin{equation}
|\Psi_{1,2}\rangle =\frac{1}{\sqrt{2\,(1+|\varepsilon |^2)}}\,
\left[ (1+\varepsilon) \,|1\rangle \pm \,(1- \varepsilon) \,
 |2\rangle \right].
\end{equation}
         In particular
\begin{equation}
H=     \left(
\begin{array}{cc}
E-\frac{i}{4}\left(\tau_q^{-1} +\tau_l^{-1}\right)
  &
\frac{1+\varepsilon}{1-\varepsilon}
\left[A-\frac{i}{4}\left(\tau_q^{-1} -\tau_l^{-1}\right)\right]
 \\  & \\
\frac{1-\varepsilon}{1+\varepsilon}
\left[A-\frac{i}{4}\left(\tau_q^{-1} -\tau_l^{-1}\right)\right]
 &
E-\frac{i}{4}\left(\tau_q^{-1} +\tau_l^{-1}\right) \\
\end{array}        \right) .
\label{HM}
\end{equation}

    Then the state $|\Psi_1 \rangle $ describes
short living particles with the life time
$ \ \tau_q \ $ and mass $E+A$.
    The state $\ |\Psi_2 \rangle $ is the state of long living particles
with life time $ \tau_l \ $  and mass $E-A$.
    Here $A$ is the arbitrary parameter $-E<A<E$  and it can be zero,
$E=M$.

   In analogy with the $K$-meson system let us take into account
transformations of the long living component into the short living
one due to the presence of baryon substance created by decays of
the short living particles.
   This process surely will depend on the density of the substance
and instead of the rough estimate in our previous paper~\cite{GrPv}
leading to some strong restrictions on the parameters of the
$CP$-violation here one obtains more realistic model.
    Let us investigate the model with the interaction which in the
basis   $\ |1 \rangle, \ |2 \rangle $  is described by the matrix
     \begin{equation}
H^d =     \left(
\begin{array}{cc}
0  & 0   \\
0  & - i \gamma \\
\end{array}        \right).
\label{Hd}
\end{equation}
               Then
\begin{equation}
\langle \Psi_1 | H^d |\Psi_2 \rangle =
\langle \Psi_2 | H^d |\Psi_1 \rangle =
i \, \frac{\gamma}{2} \, \frac{|1-\varepsilon|^2}
{1+|\varepsilon|^2}   \,.
\label{1Hd2}
\end{equation}
    which is different from the analogous expression in our
paper~\cite{GrPv} by the factor dependent on $\gamma$.
    The eigenvalues of the Hamiltonian  $H+H^d$  are
     \begin{equation}
\lambda^d_{1,2} = E - \frac{i}{4}
\left(\tau_q^{-1} + \tau_l^{-1} \right) -i\,\frac{\gamma}{2} \pm
\sqrt{ \left( A - \frac{i}{4} \left(\tau_q^{-1} - \tau_l^{-1} \right)
\right)^2 -\frac{\gamma^2}{4} } \ .
\label{lamdop}
\end{equation}

    In case  when   $\gamma \ll \tau_q^{-1}$
for the long living component one obtains
     \begin{equation}
\lambda^d_{2} \approx  E - A - \frac{i}{2}\, \tau_l^{-1}
-i\,\frac{\gamma}{2} \,,
\label{ldolg}
\end{equation}
     \begin{equation}
\| \Psi_2(t) \|{}^2 = \| \Psi_2(t_0) \|{}^2 \exp \left[
\frac{t_0 - t}{\tau_l} - \int_{t_0}^t \gamma(t)\, d t \right].
\label{P21}
\end{equation}

    The parameter $\gamma$,  describing the interaction with the
substance of the baryon medium, is evidently dependent on its state
and concentration of particles in it.
   For approximate evaluations take this parameter as
proportional to the concentration of particles:
$\gamma = \alpha\, n^{(0)}(t)$.
       Putting
$\tau_l = 2 \cdot 10^{22}$\,sec, $t \le t_U$, $a(t)=a_0 \sqrt{t}$
by~(\ref{NbM})   one obtains
     \begin{equation}
\| \Psi_2(t) \|^2 = \| \Psi_2(t_0) \|^2 \exp \left[ \alpha 2 b^{(0)}
M^{3/2}\left( \frac{1}{\sqrt{t}} - \frac{1}{\sqrt{t_0}}
\right) \right].
\label{Ptt0}
\end{equation}
      So the decay of the long living component due to this mechanism
takes place close to the time  $t_0$  which can be taken as equal
to the Compton one for $X$-particles  $t_0 \approx t_C=1/M$.
    If  $d$ -- is the part of long living particles surviving up to
the time  $t$   $\ (t_U \ge t \gg t_C)$  then from~(\ref{Ptt0})
one obtains the evaluation for the parameter~$\alpha$
     \begin{equation}
\alpha = - \ln d \,/ ( 2 b^{(0)} M^2 )\,.
\label{ald}
\end{equation}
    For  $M=10^{14}$\,Gev    and  $d=10^{-14}$   one obtains
$\alpha \approx 10^{-41}$sm${}^3$/sec.
    If  $\tau_q \sim 10^{-38} - 10^{-35} $\,sec
then the condition  $\gamma(t) \ll \tau_q^{-1} $
used in Eq.~(\ref{ldolg}) is valid for $t> t_C$.
    For this the value  $\alpha$ we have
$\gamma(t_U)\approx 10^{-47}$\,sec${}^{-1}$ $\ll \tau_l^{-1}$.
   So one can neglect this mechanism for the decay of the long living
component of $X$-particles for the modern epoch
while for early universe at $t_0 \approx t_C$ it was important.
   There is no special restriction on the parameter of
$CP$-breaking in this model.

   We supposed for simplicity the superheavy particles to be scalars.
   However one can consider them to be fermions.
   The superheavy fermions are used, for example, in some models of
neutrino mass generation (the {\it see-saw} mechanism) in
Grand Unification theories~\cite{GMRS},~\cite{Yosh}.
   Our scheme will be the same for the fermions.
    New experiments on high energetic particles in cosmic rays
surely will give us more information on their structure and origin.

\section*{Acknowledgements}

This work was supported by Min. of Education of Russia, grant E00-3-163.


\end{document}